\documentclass[sigconf]{acmart}

\renewcommand\footnotetextcopyrightpermission[1]{}
\usepackage{hyperref}
\usepackage{graphicx}
\usepackage{booktabs}
\usepackage{balance}  
\usepackage{multirow}
\usepackage{ifpdf}
\usepackage{amsmath,epsfig}
\usepackage{amssymb}
\usepackage{epstopdf}
\usepackage{subfigure}
\usepackage{caption}
\usepackage{color}
\usepackage{url}
\usepackage{mathrsfs}
\usepackage{xspace}

\newcommand{\ie}{\emph{i.e.,}\xspace}
\newcommand{\eg}{\emph{e.g.,}\xspace}

\newcommand{\eat}[1]{}

\def\EndOfProof{\nolinebreak\ \hfill\rule{1.5mm}{2.7mm}}

\begin{document}

\title{NEURON: Query Optimization Meets Natural Language Processing For Augmenting Database Education}
\author{Siyuan Liu ~~ \mbox{~~~} Sourav S Bhowmick~~~ \mbox{~~~~} Wanlu Zhang ~~~ \mbox{~~~~} Shu Wang ~~~ \mbox{~~~~} Wanyi Huang ~~~ \mbox{~~~~} Shafiq Joty}
\affiliation{%
  \institution{School of Computer Science \& Engineering, Nanyang Technological University}
  \state{Singapore}}
\email{sliu019|assourav|zh0012lu|wang1004|hu0011yi|srjoty@ntu.edu.sg}

\setcopyright{none}


\acmISBN{123-4567-24-567/08/06}


\acmPrice{15.00}

\begin{abstract}

Relational database management system (\textsc{rdbms}) is a major undergraduate course taught in many universities worldwide as part of their computer science program. A core component of such course is the design and implementation of the query optimizer in a \textsc{rdbms}. The goal of the query optimizer is to automatically identify the most efficient execution strategies for executing the declarative \textsc{sql} queries submitted by users. The query optimization process produces a \textit{query execution plan} (\textsc{qep}) which represents an execution strategy for the query. Due to the complexity of the underlying query optimizer, comprehension of a \textsc{qep} demands that a student is knowledgeable of implementation-specific issues related to the \textsc{rdbms}. In practice, this is an unrealistic assumption to make as most students are learning database technology for the first time. Hence, it is often difficult for them to comprehend the query execution strategy undertaken by a \textsc{dbms} by perusing the \textsc{qep}, hindering their learning process. In this demonstration, we present a novel system called \textsc{neuron} that facilitates natural language interaction with \textsc{qep}s to enhance its understanding. \textsc{neuron} accepts a \textsc{sql} query (which may include joins, aggregation, nesting, among other things) as input, executes it, and generates a \emph{simplified} natural language-based description (both in text and voice form) of the execution strategy deployed by the underlying \textsc{rdbms}. Furthermore, it facilitates understanding of various features related to the \textsc{qep} through a natural language-based question answering framework. We advocate that such tool, world's first of its kind, can greatly enhance students' learning of the query optimization topic.

\eat{Natural language interfaces for relational databases have been explored for several decades. Majority of the work have focused on translating natural language sentences to \textsc{sql} queries or narrating \textsc{sql} queries in natural language. Scant attention has been paid for natural language understanding of query execution plans (\textsc{qep}) of \textsc{sql} queries.  In this demonstration, we present a novel generic system called \textsc{neuron} that facilitates natural language interaction with \textsc{qep}s. \textsc{neuron} accepts a \textsc{sql} query (which may include joins, aggregation, nesting, among other things) as input, executes it, and generates a natural language-based description (both in text and voice form) of the execution strategy deployed by the underlying \textsc{rdbms}. Furthermore, it facilitates understanding of various features related to the \textsc{qep} through a natural language-based question answering framework. \textsc{neuron} can be potentially useful to database application developers in comprehending query execution strategies and to database instructors and students for pedagogical support.}
\end{abstract}

%
%
\eat{


%
%

%
%


}

\maketitle

\vspace{0ex}\section{Introduction}
Modern relational database systems (\textsc{rdbms}) are incredibly ubiquitous today -- they underlie technology used by most people every day if not every hour. As a consequence, the database system course is widely offered in major universities around the world as part of the undergraduate computer science degree program. A core component of this course is the design and implementation of the query optimizer module. Specifically, a \textsc{rdbms} employs it to automatically identify the most efficient strategies for executing the declarative \textsc{sql} queries submitted by users. The query optimization process produces a \textit{query execution plan} (\textsc{qep}) which represents an execution strategy for the query. Optimization is a mandatory process in a \textsc{rdbms} since the difference between the costs of the best execution plan, and a random choice, could be in orders of magnitude.

Unfortunately, query optimization is traditionally considered as a difficult component to fathom at an undergraduate-level database course. Given a \textsc{sql} query, a student would typically like to understand how it is executed on the underlying \textsc{rdbms} by studying the associated \textsc{qep}.  However, every commercial database vendor has its own secret sauce for the implementation of the query optimizer. Consequently, comprehension of a \textsc{qep} demands not only deep knowledge of various query optimization-related concepts but also vendor-specific implementation details.  We advocate that this is an unrealistic expectation from undergraduate students learning database systems for the first time. They may be familiar with the syntax and semantics of \textsc{sql} but not necessarily with \textsc{dbms}-specific implementation details. Consider the following example scenario.

\begin{figure}[t]
\centering
\includegraphics[width=\linewidth]{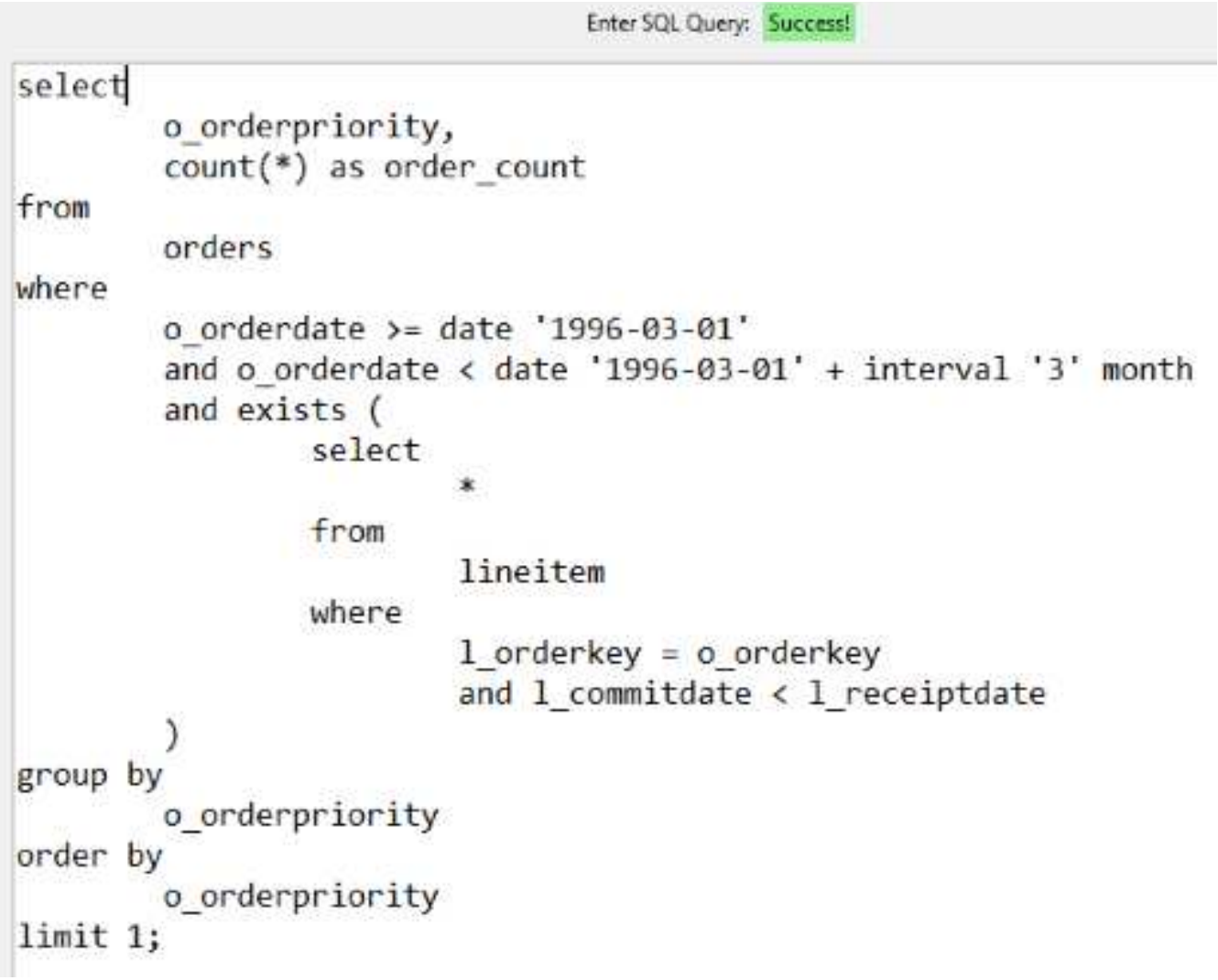}
\vspace{-1ex}\caption{Query 4 in \textsc{tpc-h} benchmark dataset.}
\label{fig:query}
\vspace{0ex}\end{figure}

\begin{figure*}[t]
\centering
\includegraphics[width=0.6\linewidth]{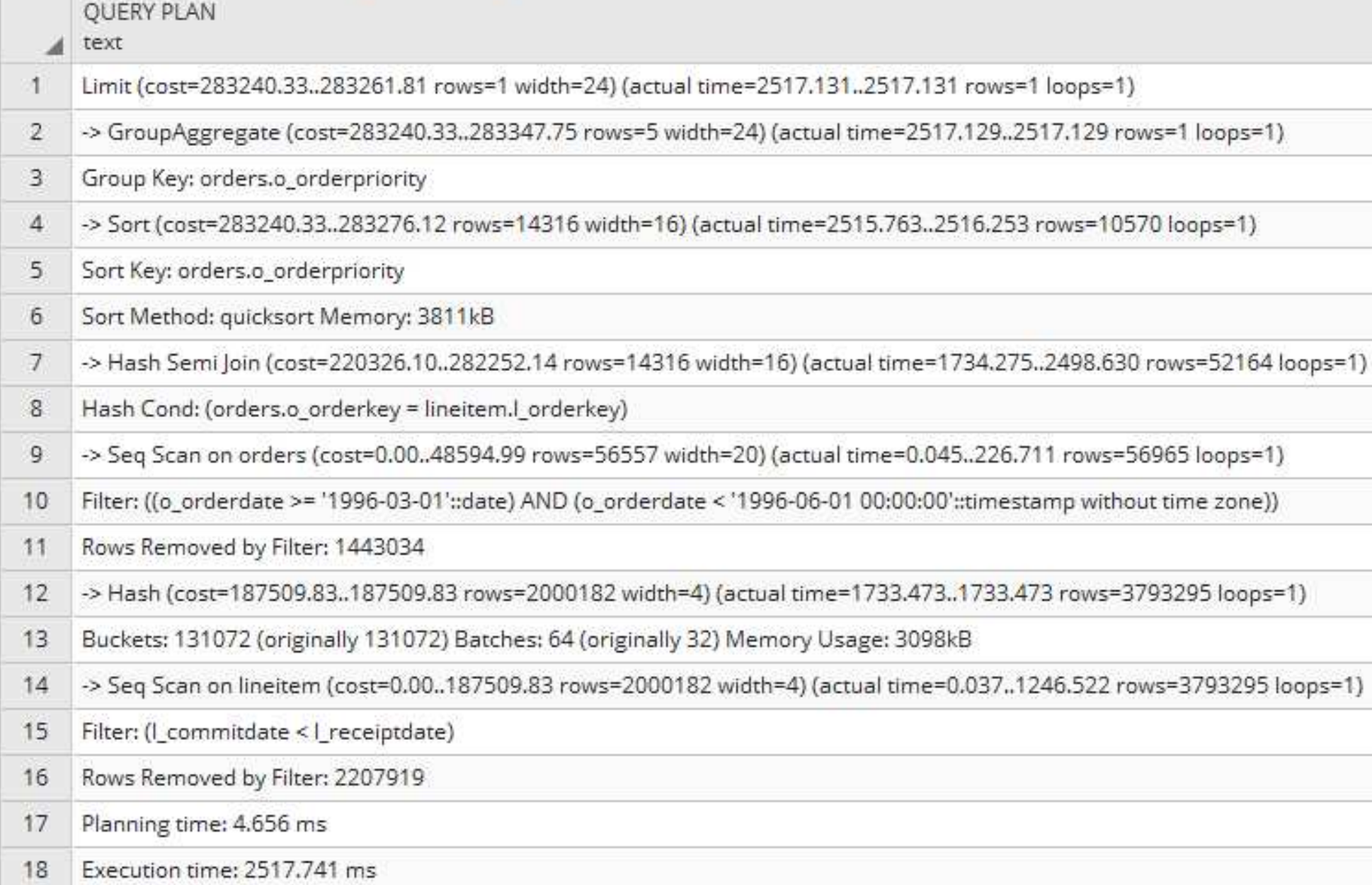}
\vspace{-1ex}\caption{A \textsc{qep} in PostgreSQL.}
\label{fig:plan1}
\vspace{0ex}\end{figure*}

\begin{figure*}[t]
\centering
\includegraphics[width=0.8\linewidth]{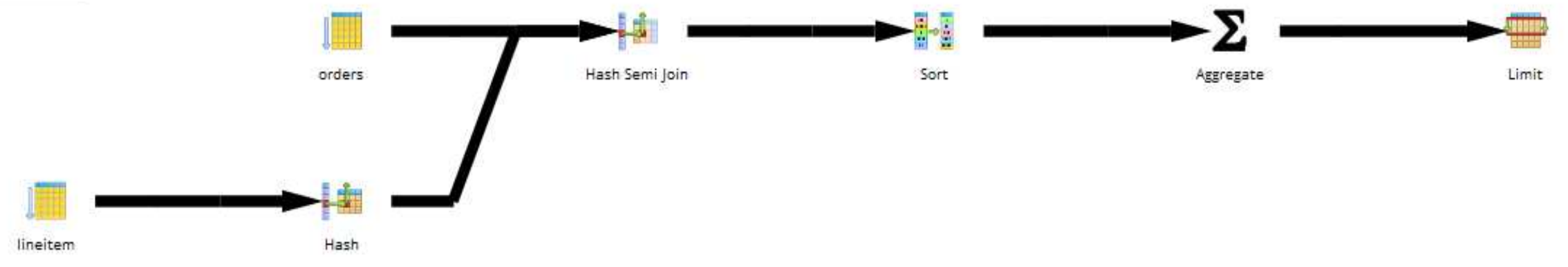}
\vspace{-1ex}\caption{Visual representation of the \textsc{qep} in PostgreSQL.}
\label{fig:plan2}
\vspace{-1ex}\end{figure*}

\begin{example}
Bob is an undergraduate sophomore student majoring in computer science in a reputed university. Currently, he is enrolled in a database course, which uses PostgreSQL 9.6 to teach fundamental concepts related to relational databases. Bob is a keen learner and is excited about learning the underlying technology behind relational database systems. Specifically, he is comfortable in writing \textsc{sql} queries and is now trying to learn about the query optimization module. To this end, he wishes to understand the \textsc{qep} of the \textsc{sql} query in Figure~\ref{fig:query} on a \textsc{tpc-h} benchmark dataset\footnote{\scriptsize \url{http://www.tpc.org}.}. Figure~\ref{fig:plan1} (partially) depicts the \textsc{qep} generated by PostgreSQL for this query. Bob observes that the textual description of the \textsc{qep} is not only verbose and lengthy but also consists of unfamiliar terms (\eg hash semijoin, bucket, width). That is, it is not concisely described in a way that can be understandable by him.

In order to have a better comprehension, he switches to the visual tree representation of the \textsc{qep} as shown in Figure~\ref{fig:plan2}. Although relatively succinct visually, it simply depicts the sequence of operators (\eg hash $\rightarrow$ hash semi join $\rightarrow$ sort $\rightarrow$ aggregate $\rightarrow$ limit) used for processing the query, hiding additional details about query execution.  In fact, Bob needs to manually delve into details associated with each node for further implementation-related information.
\EndOfProof\end{example}

Clearly, \textit{an easy and intuitive natural language-based interface can greatly enhance Bob's comprehension of \textsc{qep}s for \textsc{sql} queries}. In fact, natural language interfaces for \textsc{rdbms} have been explored by the database research community for decades~\cite{LJ14a,LJ14b,SF+16,KV+12}. Majority of these efforts have focused on translating natural language sentences to \textsc{sql} queries or narrating \textsc{sql} queries in natural language to na\"{i}ve users. Scant attention has been paid in the literature for natural language understanding of query execution plans of \textsc{sql} queries.

In this demonstration, we present a novel framework called \textsc{neuron} (\textbf{N}atural Languag\textbf{E} \textbf{U}nderstanding of Que\textbf{R}y Executi\textbf{O}n Pla\textbf{N}) for natural language interaction with \textsc{qep}s in PostgreSQL. Given the \textsc{qep} of a \textsc{sql} query, \textsc{neuron} analyzes it to automatically generate a \textit{simplified} natural language-based description (both text and voice form) of key steps undertaken by the underlying \textsc{rdbms} to execute the query. Furthermore, it supports a question-answering system that allows a user to seek answers to a variety of concepts and features associated with the \textsc{qep} in natural language.

We believe that \textsc{neuron} can be used as a tool for pedagogical support by database instructors and students. Specifically, it can facilitate understanding of various physical query plan-related concepts employed by a \textsc{rdbms} in executing \textsc{sql} queries. Furthermore, its benefit is not confined to pedagogy. It can also facilitate database application developers to understand query execution strategies employed by \textsc{sql} queries without requiring them to be knowledgeable of the syntax and semantics of \textsc{rdbms}-specific physical query plans. Note that application developers may have programming and debugging expertise to formulate declarative \textsc{sql} queries but may not necessarily possess knowledge to comprehend syntax and semantics of \textsc{rdbms}-specific \textsc{qep}s.

In this demo, we will first present a walk-through of the \textsc{neuron} tool, and explain how it provides natural language interface to understand query execution plans of modern \textsc{rdbms}. We will then show how it can be used to facilitate understanding of various concepts related to \textsc{qep}s through natural language-based question answering framework. For example, an end user may ask questions such as \textit{``What is a hash semi join?''}, \textit{``How many tuples left after Step 5?''}, and \textit{``What is the most expensive operation?''}. Finally, we will highlight how \textsc{neuron} have important implications in database education.

\vspace{0ex}\section{System Overview}
\begin{figure}
\centering
\includegraphics[width=\linewidth]{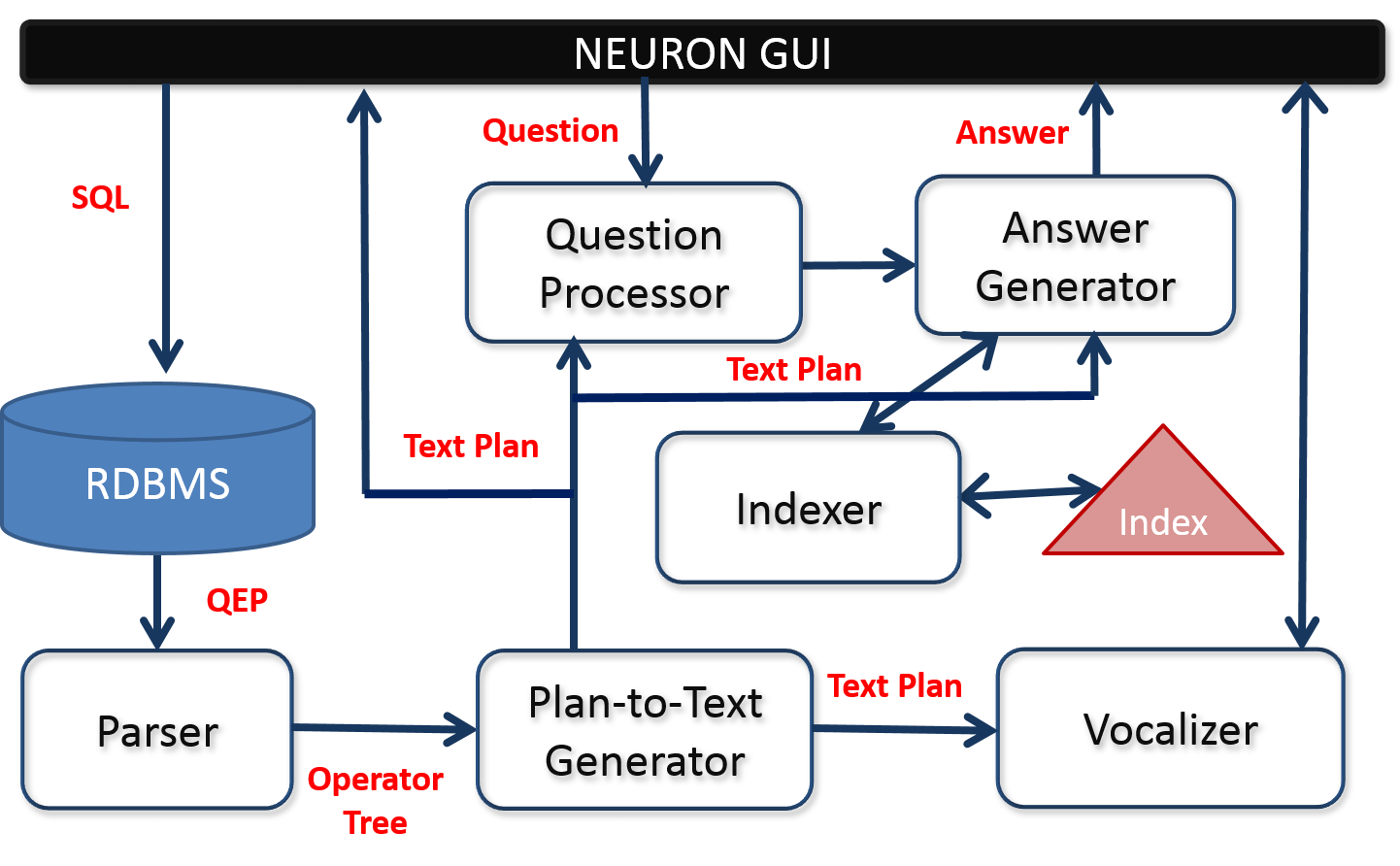}
\vspace{0ex}\caption{Architecture of \textsc{neuron}.}
\label{fig:arch}
\vspace{-1ex}\end{figure}

\textsc{neuron} is implemented using Python on top of PostgreSQL 9.6.\eat{ Note that our framework is generic in nature and can be realized on top of any commercial-strength \textsc{rdbms}.} Figure~\ref{fig:arch} depicts the architecture of \textsc{neuron} and mainly consists of the following modules.

\vspace{1ex}\noindent \textbf{\underline{The GUI module}.} Figure~\ref{fig:gui}(a) is a screenshot of the visual interface of \textsc{neuron}. It consists of five panels. Panel 1 enables a user to connect to the underlying relational database. Panel 2 shows the schema of the underlying database. A user formulates a \textsc{sql} query (which may include aggregation, nesting, joins, among other things) in textual format on this database in Panel 3. When the \texttt{Generate} button is clicked, the query is executed and the corresponding execution plan in natural language is generated and displayed in Panel 4. Note that \textsc{neuron} generates both textual as well as vocal form of the execution plan using the \textit{Plan-to-Text Generator} and \textit{Vocalizer} modules, respectively. A user can click on the \texttt{Pause} or \texttt{Replay} buttons to interact with the vocalized form of the plan. Clicking on the \texttt{View Plan} button, retrieves the original \textsc{qep} as generated by PostgreSQL. Panel 5 allows a user to pose questions related to the query execution plan in natural language by leveraging on the \textit{Question Processor} and \textit{Answer Generator} modules.

\vspace{1ex}\noindent\textbf{\underline{The Parser module}.} The goal of this module is to parse and transform the \textsc{qep} of a \textsc{sql} query into an \textit{operator tree} that will be exploited by subsequent modules. Once a user formulates and executes a \textsc{sql} query in Panel 3, it first invokes the PostgreSQL \textsc{api} (using the \textit{Psycopg} \textit{adapter}) to obtain the corresponding \textsc{qep} in \textsc{json} format. Then, the plan is parsed and an \textit{operator tree} is constructed. Specifically, each node in the operator tree contains relevant information associated with the plan such as the operator type (\eg hash join), name of the relation being processed by the node, the alias given to intermediate results (\eg subqueries),  column(s) used for grouping or sorting, the name of the index being processed by the node, subplan id generated by PostgreSQL, the condition used for searching the hash table, the filtering condition used during a join or table scan, conditions used for index-based search, and the number of rows left after an operation. These information will be subsequently utilized to generate natural language-based description of the \textsc{qep} as well as to support the question-answering framework. Note that this module ignores all information in the original \textsc{qep} that are not useful for realizing the \textsc{neuron} framework such as \textit{plan width} or whether a node is \textit{parallel aware}.

\begin{figure*}[t]
\centering
\includegraphics[width=0.9\linewidth]{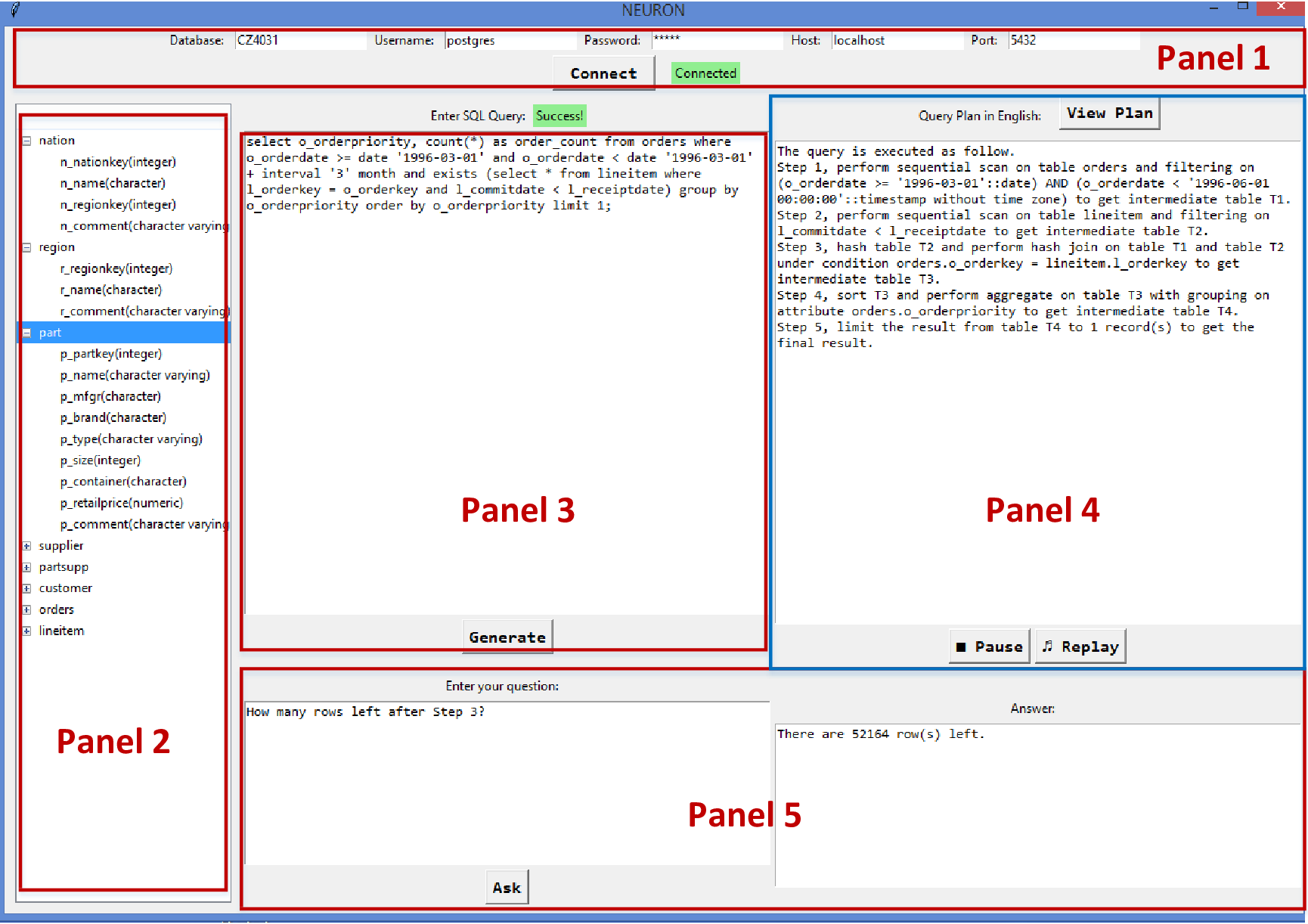}
\vspace{-1ex}\caption{Visual interface of \textsc{neuron}.}
\label{fig:gui}
\vspace{0ex}\end{figure*}

\vspace{1ex}\noindent\textbf{\underline{The Plan-to-Text Generator module}.} The objective of this module is to take the operator tree as input and generate a textual description of the \textsc{qep} represented by a sequence of steps (\eg Panel 4 in Figure~\ref{fig:gui}). At first glance, it may seem that we may simply perform a postorder traversal on the operator tree and transform the information contained in each node into natural language format. However, this na\"{i}ve approach may generate verbose description of a \textsc{qep} containing irrelevant and redundant information. This is because some nodes in an operator tree may not carry meaningful information as far as textual description of a \textsc{qep} is concerned. For instance, the node \textsf{Result} is used in PostgreSQL to represent intermediate relation for storing temporary results. Although it is an important step for executing a query, it is unnecessary to show it as an individual step in our output. Hence, this module first removes \textsf{Result} nodes from the operator tree.

The modified operator tree contains now two categories of nodes, namely \textit{critical} and \textit{non-critical} nodes. The former nodes represent important operations (\eg hash join, sort) in a \textsc{qep} and may contain a large amount of information. On the hand, the latter nodes are located near critical nodes (\eg parent, child) but do not carry important information on its own in comparison to the critical ones. Hence, we reduce the modified operator tree further by \textit{merging} the non-critical nodes with corresponding critical nodes. Some examples of such merge operation are as follows.
\begin{itemize}
        \item The \textsf{Hash Join} node and its child \textsf{Hash} are merged.
        \item The \textsf{Merge Join} node and its children \textsf{Sort} are merged.
        \item The \textsf{Bitmap Heap Scan} node and its child \textsf{Bitmap Index Scan} are merged.
        \item The \textsf{Aggregate} node and its child \textsf{Sort} are merged.
        \item The \textsf{Unique} node and its child \textsf{Sort} are merged.
        \end{itemize}

An important issue to address while generating a natural language representation of a \textsc{qep} is the handling of subqueries in a \textsc{sql} query. PostgreSQL creates a corresponding subplan for each subquery in the \textsc{qep} whose return value can be referred to from other parts of the plan. It assigns a temporary name to this subplan for future referral. However, this name should not appear in the natural language representation of the \textsc{qep}. Thus, we use a dictionary to keep track of the subplan names and their corresponding relation names so that when other steps mention the output of the subquery, the referred name will be replaced by the corresponding relation name(s).

Based on the aforementioned strategies, this module generates the natural language representation of a \textsc{qep} from the reduced operator tree as follows. It traverses the tree in postorder fashion to generate a sequence of steps (identified by \textit{step id}) describing the \textsc{qep}. Each node in the reduced operator tree generates a step and each step is represented as a  text description of the node's content based on its type. Specifically, we leverage different \textit{natural language templates} for different node types to generate meaningful statements. In this context, each intermediate result is assigned an identifier. This allows a clear reference from a parent operator to its children's result without any ambiguity. Filter and join conditions are parsed and converted to human readable natural language representation. For example, an \textsf{Index Scan} node is converted to the following step: ``\textit{Perform index scan on table \textsf{X} (and filtering on \textsf{X.b = 1}) to get intermediate table \textsf{A}}''. Figure~\ref{fig:gui} depicts an example of the output of this module (in Panel 4) for the \textsc{qep} in Figure~\ref{fig:plan1}.

It is worth noting that the textual description of the \textsc{qep} generated by this module is richer in implementation-specific information of a query compared to textual narrative generated from a declarative \textsc{sql} query by tools like Logos~\cite{KV+12}. This is because execution-specific details (\eg type of join, type of scan) of a \textsc{sql} query cannot be simply gleaned from its declarative statement.

\vspace{1ex}\noindent\textbf{\underline{The Vocalizer module}.} The goal of this module is to vocalize the natural language description of the \textsc{qep} generated by the \textit{Plan-to-Text Generator} module. Specifically, the text to speech conversion is performed utilizing Google's Text-to-Speech (\textsc{gtts}) \textsc{api} and played using the \textit{Pygame} package (\url{https://www.pygame.org/}).

\vspace{1ex}\noindent\textbf{\underline{The Indexer module}.} This module is exploited by the question-answering (\textsc{qa}) framework of \textsc{neuron}. The \textsc{qa} subsystem accepts a user query as input and returns an answer as output (Panel 5). Note that not all queries related to a \textsc{qep} can be answered by analyzing the \textsc{qep}. For example, \textit{``what is a bitmap heap scan?''} cannot be answered simply by analyzing the \textsc{qep}. To address this challenge, this module first extracts definitions of \textsc{sql} keywords and query plan operators from relevant Web sources\footnote{\scriptsize \url{https://www.postgresql.org/docs/10/static/sql-commands.html}, \url{http://use-the-index-luke.com/sql/explain-plan/postgresql/operations}, \url{https://www.postgresql.org/message-id/12553.1135634231@sss.pgh.pa.us}} as well as comments associated with source code of PostgreSQL\footnote{\scriptsize  \url{https://github.com/postgres/postgres/blob/master/src/include/nodes/plannodes.h}}. Then a set of documents containing these definitions are indexed using an inverted index (we use the \textit{Whoosh} Python library (\url{https://pypi.python.org/pypi/Whoosh/})). where each document contains the definition of a single \textsc{sql} keyword or query operator. The words in documents are lemmatized and stop words are removed during this process.

\begin{figure*}[t]
\centering
\includegraphics[width=0.8\linewidth]{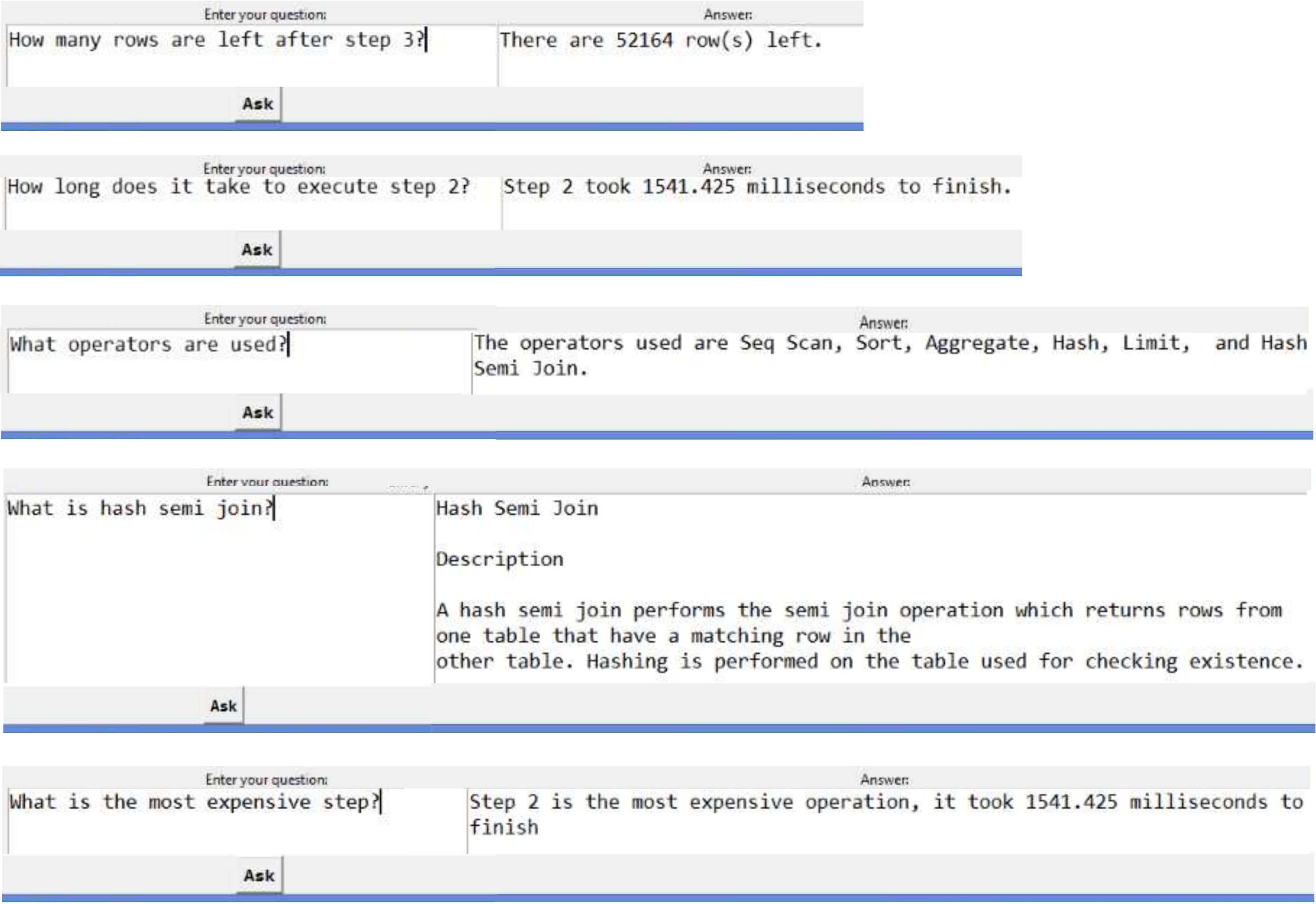}
\vspace{-1ex}\caption{Question-answering subsystem of \textsc{neuron}.}
\label{fig:answer}
\vspace{0ex}\end{figure*}

\eat{\begin{figure*}[t]
\centering
\includegraphics[width=\linewidth, height=6.7cm]{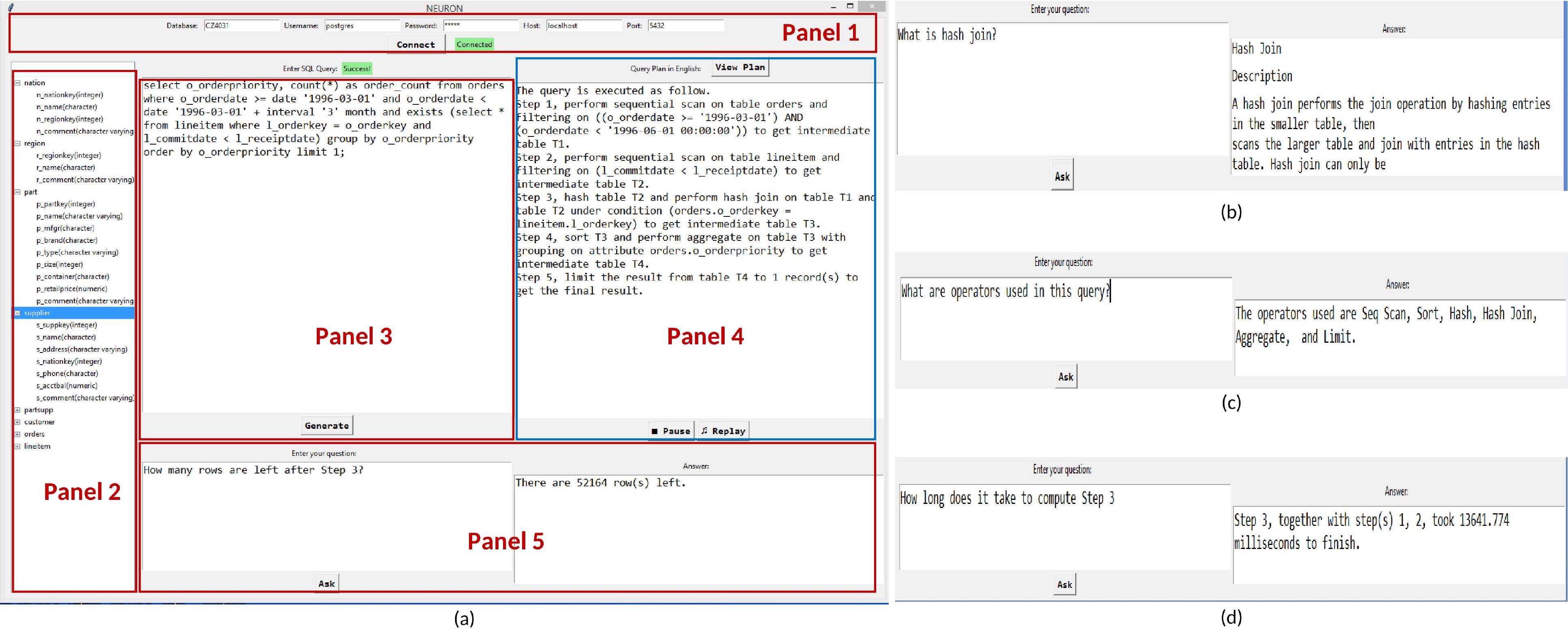}
\vspace{-2ex}\caption{\textsc{gui} of \textsc{neuron} and question-answering subsystem.}
\label{fig:gui}
\vspace{-2ex}\end{figure*}}

\vspace{1ex}\noindent\textbf{\underline{The Question Processor module}.} Once a user enters a question related to the \textsc{qep} through Panel 5, the goal of this module is to \textit{classify} the question, and extract the part-of-speech (\textsc{pos}) tags and keywords in the question. Consequently, it consists of three submodules, namely, the \textit{question classifier}, the \textit{part-of-speech (POS) tagger}, and the \textit{keyword extractor} submodules. We elaborate on them in turn.

\textit{\underline{The Question Classifier submodule}.} The current implementation of \textsc{neuron} supports five categories of questions: (a) definitions of various \textsc{sql} keywords and query plan operators; (b) the number of tuples generated at a specific step; (c) list of operators used to evaluate the query; (d) the amount of time taken by specific step(s) in the \textsc{qep}; and (e) finding the \textit{dominant} (\ie most expensive) operator in the \textsc{qep}.  Hence, given a user's question, its category needs to be identified first before it can be answered. The goal of this submodule is to classify a user's question in one of these five categories. To this end, it adopts the Naive Bayes, a learning-based classification method. A set of training questions were prepared manually together with their true categories. As there are five categories, it is not necessary to generate a very large number of training questions (we use 67 questions for training). The features used for the classification is the bag of words. Our experiments show that this strategy is effective in classifying different questions accurately.

Given a user's question, the bag of words feature is generated for the question and the question category is obtained from the classifier.

\textit{\underline{The Part-of-speech (\textsc{POS}) Tagger submodule}.} This submodule extracts the part-of-speech (\textsc{pos}) tags in a question\footnote{\scriptsize Our implementation uses the TextBlob Python library (\url{https://pypi.python.org/pypi/textblob}) using the Penn Treebank corpus. It is a high-level NLP toolkit in Python built on top of NLTK.}. \textsc{pos} tags are used to find the \textit{step id} (\ie \textit{id} of a step in Panel 4) inside a question related to Categories (b) and (d).

\textit{\underline{The Keyword extractor submodule}.} To answer questions related to Category (a), it is paramount to identify the keywords in the question so that we know what is being asked. This submodule extracts the keywords by first removing stop words. The list of English stop words is obtained from the \textsc{nltk} Python library (\url{http://www.nltk.org/}). The word \textsf{only} is excluded as it is one of the keywords for query operators (\eg \textsf{Index Only Scan}). The remaining words are lemmatized and duplicate words are eliminated.

\vspace{1ex}\noindent\textbf{\underline{The Answer Generator module}.} Given a question, the \textit{Question Processor} module identifies its category, relevant keywords and step id. The \textit{Answer Generator} module aims to retrieve the correct answer based on the question category. As there are five categories of questions, different submodules are designed to handle them.

\textit{\underline{The Concept Definition submodule}.} If the question belongs to Category (a) then it uses keywords extracted from it to retrieve the relevant document containing the definition using the index.

\textit{\underline{The Row Count submodule}.} To answer questions regarding the number of rows after a certain step (Category (b)), the \textit{step id} must be supplied in the question. Note that questions in the form of ``\textit{number of rows left after joining relations \textsf{A} and \textsf{B}}'' (\ie without step id) are not supported. This is because it is possible that two or more joins on the same relations but different columns may be performed in a single query, leading to ambiguity.

The submodule extracts the step id by finding word with the \textsc{pos} tag \textsf{CD} (cardinal number) in the question. After that, the operator tree is traversed to find the node the step id belongs to. The number of rows is retrieved from the \textsf{Actual Rows} element associated with the query plan node.

\textit{\underline{The Operator List submodule}.} To retrieve the operators used in a \textsc{qep} (Category (c)), the operator tree is traversed. Duplicate operators are removed and the final list is returned to Panel 5.

\textit{\underline{The Total Time submodule}.} To answer questions regarding Category (d), similar to Category (b) questions, the \textit{step id} must be supplied in the question. It traverses the operator tree to retrieve the total time of a specific step, which is calculated based on the \textsf{Actual Total Time} element of the node itself and its children. The returned answer includes the actual time spent on the queried step.

\textit{\underline{The Dominant Operator submodule}.}  To find the most expensive operator in the \textsc{qep} (Category (e)), \textsc{neuron} computes the total time taken by each operator and returns the one with longest time.

Note that the answers are formatted using natural language templates to generate meaningful statements. Figure~\ref{fig:answer} depicts example screenshots of several types of questions supported by the \textsc{qa} subsystem of \textsc{neuron}.

\vspace{0ex}\section{Related Systems and Novelty}
Query optimizers have been extensively studied since the inception of relational databases. Several interesting features of query optimizers have been demonstrated in conference venues as well. For example, \textsc{picasso}~\cite{picasso} is a visualization tool for graphically profiling and analyzing the behavior of database query optimizers. QE3D~\cite{SDB15} is another query plan visualization tool that provides holistic view of distributed query plans executed by the \textsc{sap hana} database management system. Stethoscope~\cite{GK12} is an interactive visual tool to analyze plans for a columnar database.  However, to the best of our knowledge, there has been no prior work on natural language understanding of query plans\eat{ with regard to real-world query optimizers}.

Natural language interfaces to databases have been studied for several decades. Such interfaces enable users easy access to data, without the need to learn a complex query languages, such as \textsc{sql}. Specifically, there have been natural language interfaces for relational databases~\cite{LJ14a,LJ14b,SF+16,PEK03,BH+18}, video databases~\cite{ECC08}, \textsc{xml}~\cite{LC+07}, and graph-structured data~\cite{ZC+17}. Given a logically complex English language sentence as query input, the goal of majority of these work is to translate them to the underlying query language such as \textsc{sql}. On the other hand, frameworks such as Logos~\cite{KV+12} explain \textsc{sql} queries to naive users using natural language. \textsc{neuron} compliments these efforts by providing a natural language explanation of the query execution plan of a given \textsc{sql} query. It further supports a natural language-based question answering framework that enables users to ask questions related to the plan.

\vspace{0ex}\section{Demonstration Objectives}
Our demonstration will be loaded with \textsc{tpc-h} benchmark (we use the \textsc{tpc-h} v2.17.3 at \url{http://www.tpc.org/tpc_documents_current_versions/current_specifications.asp}) and \textsc{dblp} datasets. For \textsc{dblp}, we download the \textsc{xml} snapshot of the data and then store them in 10 relations. Example \textsc{sql} queries on these datasets will be presented. Users can also write their own ad-hoc queries through our \textsc{gui}.

One of the key objectives of the demo is to enable the audience to interactively experience the benefits of this novel natural language interface for query execution plans in real-time. The audience will be requested to formulate a \textsc{sql} query or select one from the list of benchmark queries using the \textsc{neuron} \textsc{gui}. Upon execution of the query, one will be able to view as well as hear the natural language description of the \textsc{qep} (through the \textit{Plan-to-Text Generator} and \textit{Vocalizer} modules). She may pause and replay the natural language description as she wishes. By clicking on the \texttt{View Plan} button, one can view the original \textsc{qep} generated by PostgreSQL and appreciate the difficulty in perusing and comprehending the details of the plan, highlighting the benefits of natural language interaction brought by \textsc{neuron}. Lastly, the audience can pose the aforementioned types of questions related to a \textsc{qep} through the \textsc{neuron} \textsc{gui} and get accurate answers in real-time. Such \textsc{qa} session aims to facilitate further natural language-based clarification regarding the execution strategy deployed by the underlying query engine.

\vspace{0ex}\section{Illustration of Example Use Case}
A short video to illustrate the aforementioned features of \textsc{neuron} using an example use case on \textsc{tpc-h} benchmark data is available at \url{https://youtu.be/wRIWuYbU2F0}. Specifically, it emphasizes the ease with which a user can interaction with \textsc{neuron}, natural language description of the \textsc{qep} of an example query, and interactive question-answering sessions demonstrating the five categories of questions related to the \textsc{qep}.





\vspace{0ex}\begin{thebibliography}{10} \small

\eat{\bibitem{DH16} A. Dutt, J. R. Haritsa. Plan Bouquets: A Fragrant Approach to Robust Query Processing. \textit{ACM Trans. Database Syst.}, 41(2), 2016.}

\bibitem{BH+18} F. Basik, B. H\"{a}ttasch, et al. DBPal: A Learned NL-Interface for Databases. \textit{In SIGMOD}, 2018.

\bibitem{ECC08} G. Erozel, N. K. Cicekli, I. Cicekli.  Natural language querying for video databases. \textit{Inf. Sci.}, 178(12), 2008.

\bibitem{GK12} M. Gawade, M. L. Kersten. Stethoscope: A platform for interactive visual analysis of query execution plans. \textit{In PVLDB}, 5(12), 2012.

\bibitem{picasso} J. R. Haritsa.  The Picasso Database Query Optimizer Visualizer. \textit{In PVLDB}, 3(2), 2010.

\bibitem{KV+12} A. Kokkalis, P. Vagenas, A. Zervakis, A. Simitsis, G. Koutrika, Y. E. Ioannidis. Logos: a system for
translating queries into narratives. \textit{In SIGMOD}, 2012.

\bibitem{LJ14a} F. Li, H. V. Jagadish. NaLIR: an interactive natural language interface for querying relational databases. \textit{In SIGMOD}, 2014.

\bibitem{LJ14b} F. Li, H. V. Jagadish. Constructing an Interactive Natural Language Interface for Relational Databases. \textit{PVLDB}, 8(1), 2014.

\bibitem{LC+07} Y. Li, I. Chaudhuri, H. Yang, S. P. Singh, H. V. Jagadish.
DaNaLIX: a domain-adaptive natural language interface for querying XML. \textit{In SIGMOD}, 2007.


\bibitem{LL+17}	J. Lin, Y. Liu, J. Guo, J. Cleland-Huang, W. Goss, W. Liu, S. Lohar, N. Monaikul, A. Rasin.
TiQi: a natural language interface for querying software project data. \textit{In ASE}, 2017.

\bibitem{PEK03} A.-M. Popescu, O. Etzioni, H. A. Kautz. Towards a theory of natural language interfaces to databases. \textit{In IUI}, 2003.

\bibitem{SF+16} D. Saha, A. Floratou, et al.
ATHENA: An Ontology-Driven System for Natural Language Querying over Relational Data Stores. \textit{In PVLDB}, 9(12), 2016.

\bibitem{SDB15} D. Scheibli, C. Dinse, A. Boehm. QE3D: Interactive Visualization and Exploration of Complex, Distributed Query Plans. \textit{In SIGMOD}, 2015.

\bibitem{ZC+17} W. Zheng, H. Cheng, L. Zou, J. X. Yu, K. Zhao. Natural Language Question/Answering: Let Users Talk With The Knowledge Graph. \textit{In CIKM}, 2017.

\eat{\bibitem{ZP+17} J. Zhu, N. Potti, S. Saurabh, J. M. Patel. Looking Ahead Makes Query Plans Robust. \textit{In PVLDB}, 10(8), 2017.}

\end{thebibliography}
\end{document}